\documentclass[12pt]{article}
\usepackage[utf8]{inputenc}
\usepackage{Preambles/preamble}
\RequirePackage{doi}
\usepackage{hyperref}

\begin{document}

\submitdate{\today}
\maketitle

\thispagestyle{empty}

\newpage
\hrule height .05in
\vspace{.2in}

\begin{center}
{\Large \textbf{Harvard Undergraduate Survey on Generative AI}}

\hrulefill
\vspace{.2in}

\renewcommand{\thefootnote}{\fnsymbol{footnote}}
\makeatletter
\textbf{Shikoh Hirabayashi\footnote{Harvard Undergraduate Association} \qquad Rishab Jain \qquad Nikola Jurkovi\'c\footnote{AI Safety Student Team.\\Correspondence to \textit{gabrielwu@college.harvard.edu}. Authors listed in alphabetical order. This report was commissioned by the Harvard Undergraduate Association and written by Harvard College students, but it is not otherwise affiliated with Harvard University.} \qquad Gabriel Wu\footnotemark[2]}
\end{center}
\makeatother
\renewcommand{\thefootnote}{\arabic{footnote}}
\setcounter{footnote}{0}

\vspace{.5in}
\begin{abstract}
\normalsize
\noindent How has generative AI impacted the experiences of college students? We study the influence of AI on the study habits, class choices, and career prospects of Harvard undergraduates ($n=326$), finding that almost 90\% of students use generative AI. For roughly 25\% of these students, AI has begun to substitute for attending office hours and completing required readings. Half of students are concerned that AI will negatively impact their job prospects, and over half of students wish that Harvard had more classes on the future impacts of AI. We also investigate students' outlook on the broader social implications of AI, finding that half of students are worried that AI will increase economic inequality, and 40\% believe that extinction risk from AI should be treated as a global priority with the same urgency as pandemics and nuclear war. Around half of students who have taken a class on AI expect AI to exceed human capabilities on almost all tasks within 30 years. We make some recommendations to the Harvard community in light of these results.
\end{abstract}


\newpage
\tableofcontents

\newpage
\section{Introduction} \label{sec:introduction}

\subsection{The Rise of Generative AI: A Student Perspective}
In recent years, generative AI technology --- AI systems trained to produce original text, images, and videos --- has advanced rapidly. Today, the most advanced chatbots powered by GPT-4 and Claude 3 can generate stunning images in any art style, perform basic autonomous tasks like browsing the web \cite{kinniment2024evaluating}, and consistently outperform most humans on standardized tests.\footnote{Claude 3 has an accuracy of 60\% on GPQA, a dataset of extremely difficult science questions \cite{Claude3}. For reference, human PhD students score 65\% on questions in their field of expertise, and only 34\% on questions from a different field --- even with access to Google.} A recent survey of thousands of AI experts estimated that there is a 50\% chance of AI outperforming humans in all tasks by 2047 \cite{grace2024thousands}.

This rapid progress has raised many questions about how AI will affect education. How should pedagogy change when a chatbot can generate an essay in seconds or walk a student through a problem set as effectively as any human tutor? Will the skills that we teach in the classroom be relevant to graduates entering an increasingly AI-centric economy? \cite{eloundou2023gpts}

If this trend of AI progress continues, generative AI will also have a large impact on the experience of students more broadly. AI may affect mental health \cite{ettman2023mentalhealth} and economic inequality among students. Moreover, AI raises important societal questions about responsible deployment and the future role of humans in an AI-transformed economy.

In an attempt to shed light on current student perspectives surrounding these questions, the Harvard Undergraduate Association\footnote{In collaboration with the AI Safety Student Team, the Harvard Undergraduate Open Data Project, and the Harvard Undergraduate Machine Intelligence Community.} commissioned the inaugural ``Survey on Generative AI'' this April. In this report, we present key findings: use of generative AI among students is nearly pervasive, income levels shape the impact of AI usage, and students have a wide range of opinions about AI's future, reflecting both optimism and concern.

\subsection{Purpose of this report}
The primary goal of this report is to investigate the ways in which generative AI has affected student experiences at Harvard College. The survey questions cover a wide variety of AI-related topics, including the influence of AI on students' study habits, class choices, and career prospects. It also explores students' hopes and concerns regarding increased economic inequality, concentration of power, and risks of human extinction related to AI.

A second goal of this report is to inform AI-relevant decisions made by Harvard University. In Section \ref{sec:recommendations}, we interpret key findings from the survey to offer recommendations to the Harvard community. These relate to course offerings, mental health resources, career support, and more generally, preparation for societal-scale disruption from generative AI. While these recommendations are primarily addressed to the Harvard administration, we expect them to generalize to other academic institutions as well.

Finally, a third purpose of this report is to serve as an example for future surveys on AI. To our knowledge, this is the first public report of its kind conducted on a college campus. We hope it paves the way for future work at universities across the world.

\section{Results}\label{Results}

We collected responses through a survey sent out to all Harvard undergraduates. We received responses from 326 students, which were filtered down to 273 using a reading comprehension check. See Appendix \ref{app:survey_details} for details.

The key results are split across four subsections: general trends in generative AI usage (\S \ref{sec:ai_use}), the impact of AI on students' study habits (\S \ref{sec:study_habits}), the impact of AI on students' course selection and career plans (\S \ref{sec:course_and_career}), and broader concerns about the societal impacts of AI (\S \ref{sec:broader_concerns}). For brevity we only include the most important figures in this section; more results and figures can be found in Appendix \ref{app:more_figures}.

Our sample is representative in terms of gender and class year, although Asians and Computer Science concentrators are slightly overrepresented. More analysis on our sample's demographic representativeness can be found in Appendix \ref{app:representativeness}.

\subsection{Overall AI use} \label{sec:ai_use}

\begin{figure}
    \centering
    \includegraphics[width= 6in]{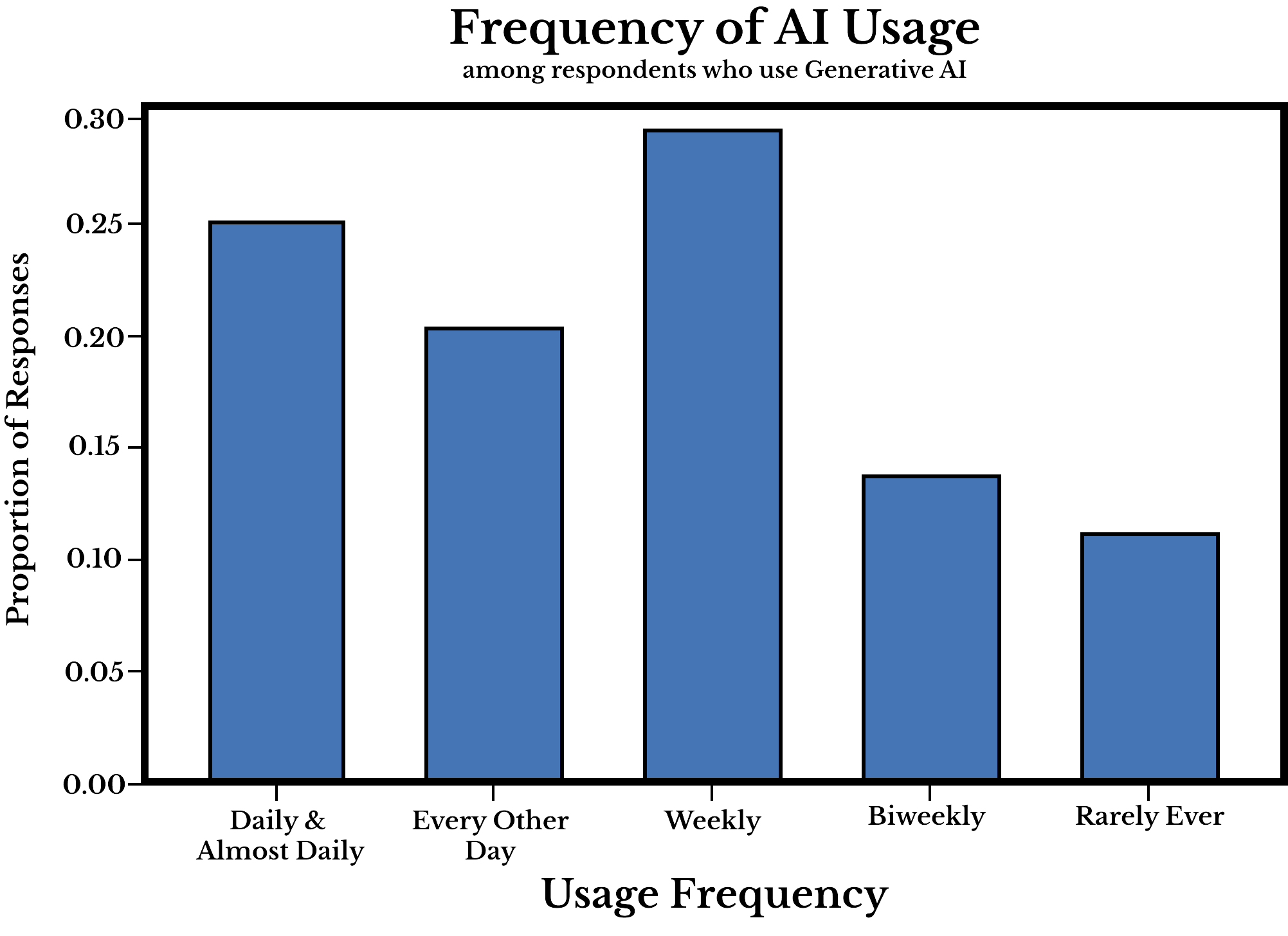}
    \caption{Most Harvard students who use generative AI report using it at least once a week.}
    \label{fig:frequency_of_usage}
\end{figure}

A large majority of respondents (87.5\%) reported using generative AI.\footnote{The question was phrased as: ``Do you ever use generative AI products? These include chatbots, image generators, or AI music generators.''} Among respondents who use generative AI, most reported using it at least once a week, and almost half use it at least every other day (Figure \ref{fig:frequency_of_usage}).

\begin{figure}
    \centering
    \includegraphics[width=0.75\linewidth]{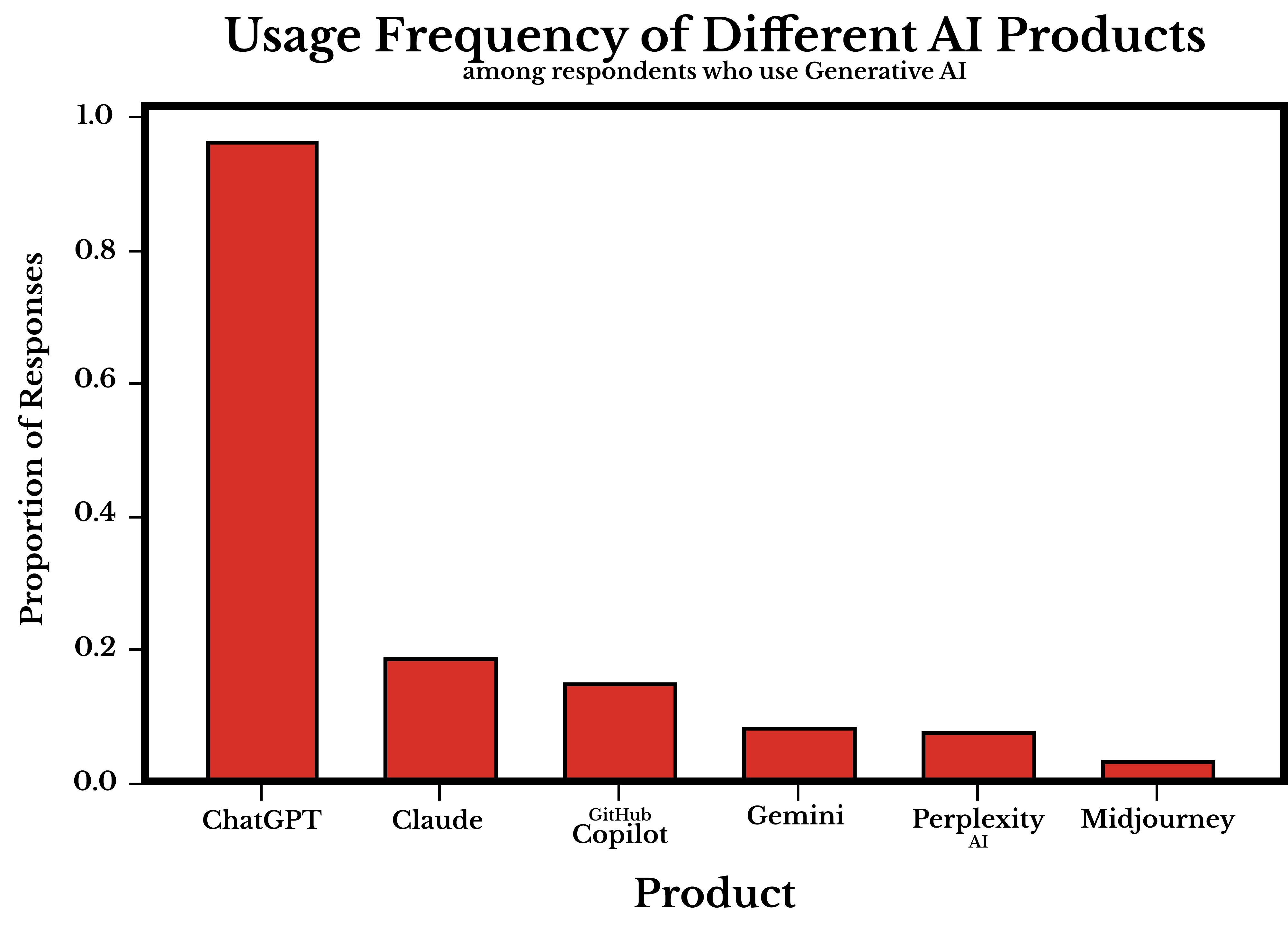}
    \caption{The proportion of respondents who use generative AI that report using each of six popular AI products. OpenAI's ChatGPT is by far the most widely used product.}
    \label{fig:models_compared}
\end{figure}

There is not much variety among the types of AI products that students use. ChatGPT is extremely popular --- over 95\% of AI users report using it --- while Claude (Anthropic's language model) and GitHub Copilot (a programming assistant) are each used by around 20\% of AI users (Figure \ref{fig:models_compared}).

\begin{figure}
    \centering
    \includegraphics[width=\linewidth]{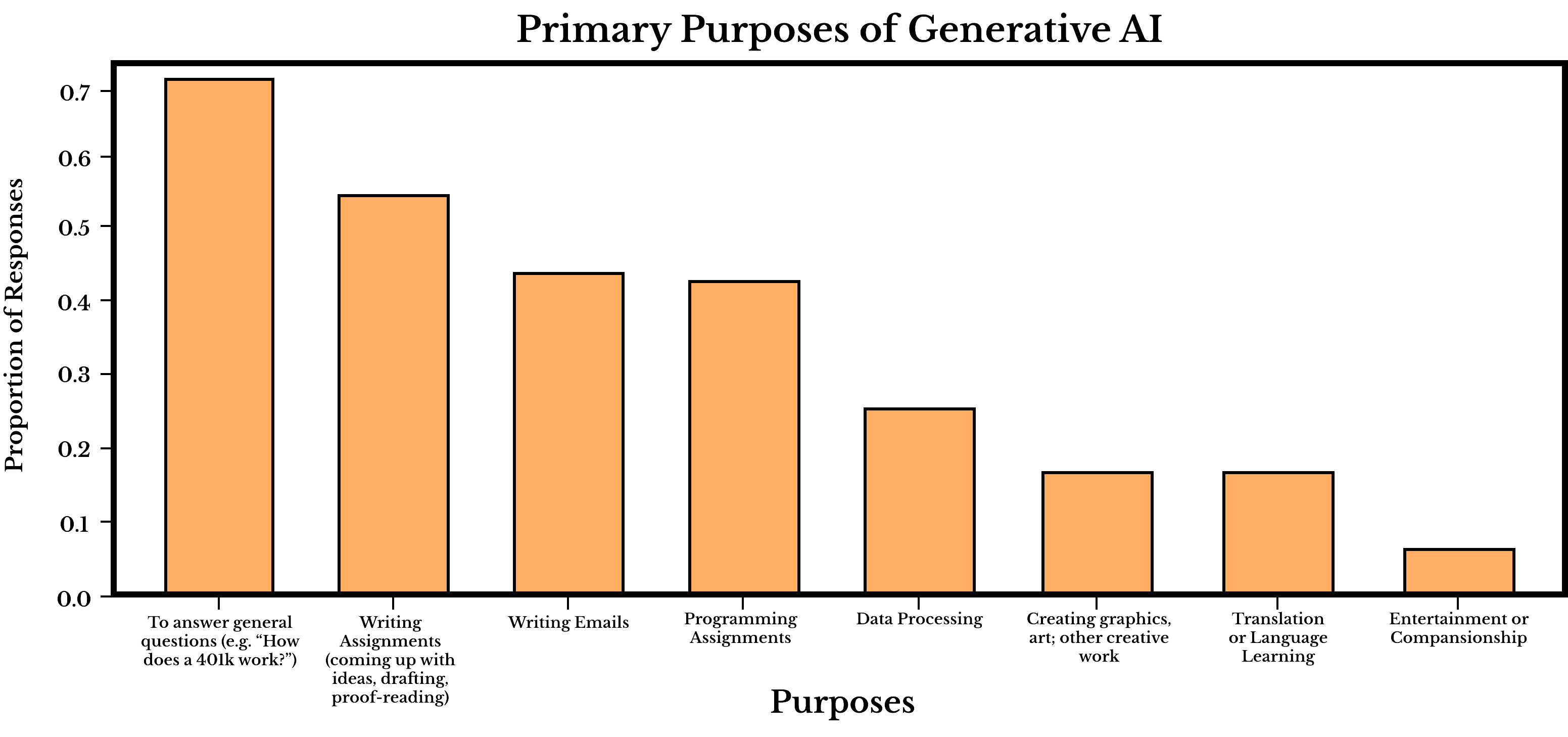}
    \caption{Students use generative AI for a wide variety of purposes, including answering general questions and help writing essays, emails, and code.}
    \label{fig:types_of_use}
\end{figure}

However, there is wide variety in the ways students use AI (Figure \ref{fig:types_of_use}). The most common use of generative AI is to answer general questions, like ``How does a 401k work?''. In fact, for around a third of students, AI is replacing the role of traditional informational sources like Wikipedia and Google search (Figure \ref{fig:study_habits}). Other common uses of AI are: help with writing assignments (e.g. coming up with ideas, drafting, proof-reading), writing emails, and helping with programming assignments and data processing.

How common is it for students to pay money for AI products? 30\% of respondents who use generative AI report paying for premium AI subscriptions (Figure \ref{fig:spending}). It is likely that most of these subscriptions are to ChatGPT Plus, which currently costs \$20 per month. Paying money makes a difference: students who spend money on AI report getting more benefit from it, and rely less on University resources and traditional search engines (Figure \ref{fig:habits_by_spending}). We also find that students coming from lower socioeconomic backgrounds are significantly less likely to spend money on AI (Figure \ref{fig:spending_by_finaid}). See Appendix \ref{app:spending} for more details and discussion on the impact of paid subscription plans.

\subsection{Impact on study habits} \label{sec:study_habits}

\begin{figure}
    \centering
    \includegraphics[width=\linewidth]{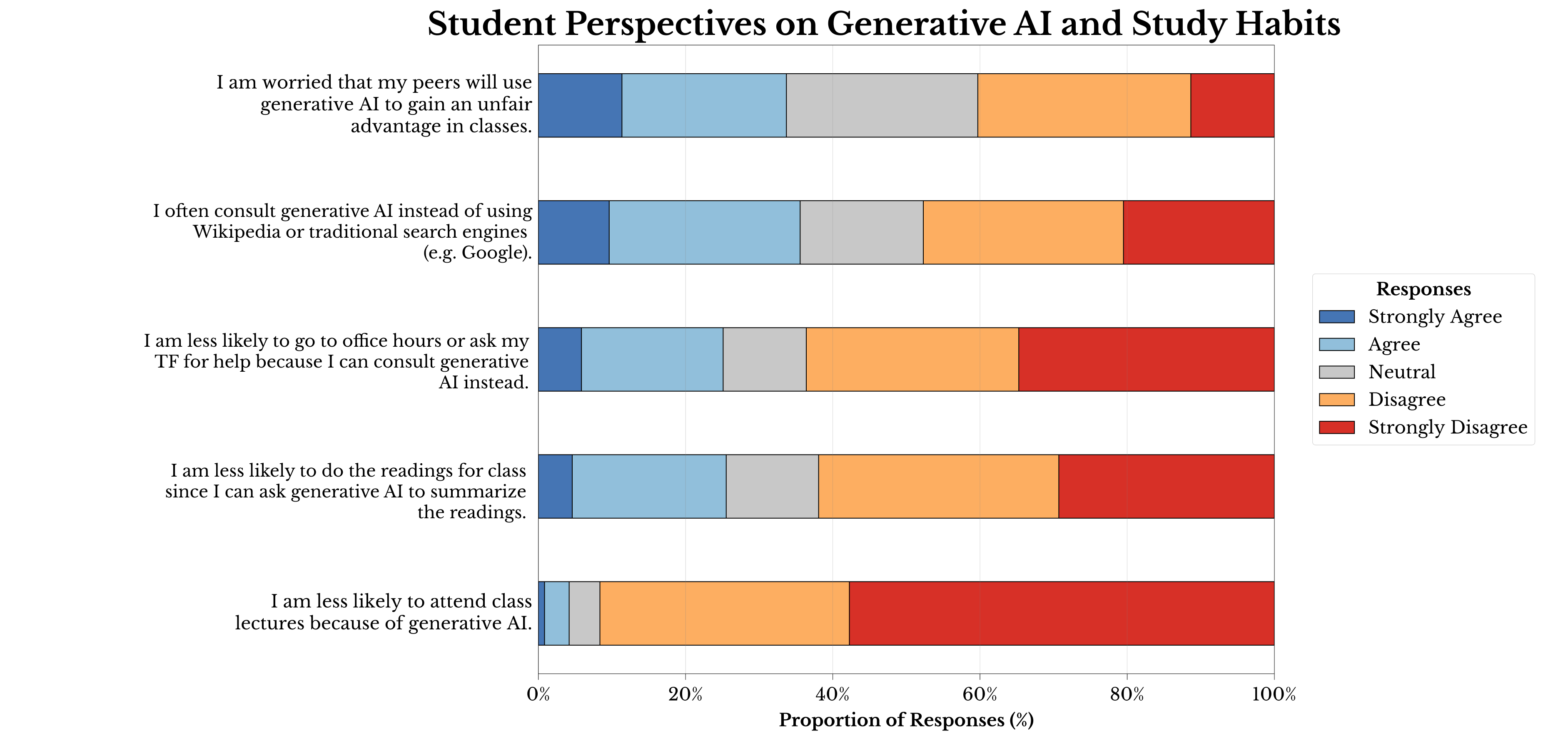}
    \caption{Impact of AI use on study habits, among students who use generative AI (with the exception of ``I am worried that my peers will use generative AI to gain an unfair advantage in classes,'' which was asked to all respondents).}
    \label{fig:study_habits}
\end{figure}

Figure \ref{fig:study_habits} shows how students' study habits have been shaped by generative AI. Around 25\% of students who use generative AI find themselves going to office hours, asking course staff for help, and completing required readings less often because of the availability of generative AI.\footnote{In a free-response answer, one student remarked that ``the questions about if i rather use Ai to get help instead of going to office hours hit a a little too deep.''} However, very few students report going to class lectures themselves less because of AI.

Around 35\% of students are worried their peers will use generative AI to gain an unfair academic advantage in class, suggesting that Harvard should be thoughtful and explicit about creating enforceable rules around AI use. In particular, this concern revolves around \textit{enforceability} as opposed to simply making the rules clear: only 4\% of respondents disagreed with the sentence ``I understand the rules regarding the use of generative AI in my classes,'' and \textit{nobody} strongly disagreed.

\subsection{Impact on course selection and career plans} \label{sec:course_and_career}

\begin{figure}
    \centering
    \includegraphics[width=\linewidth]{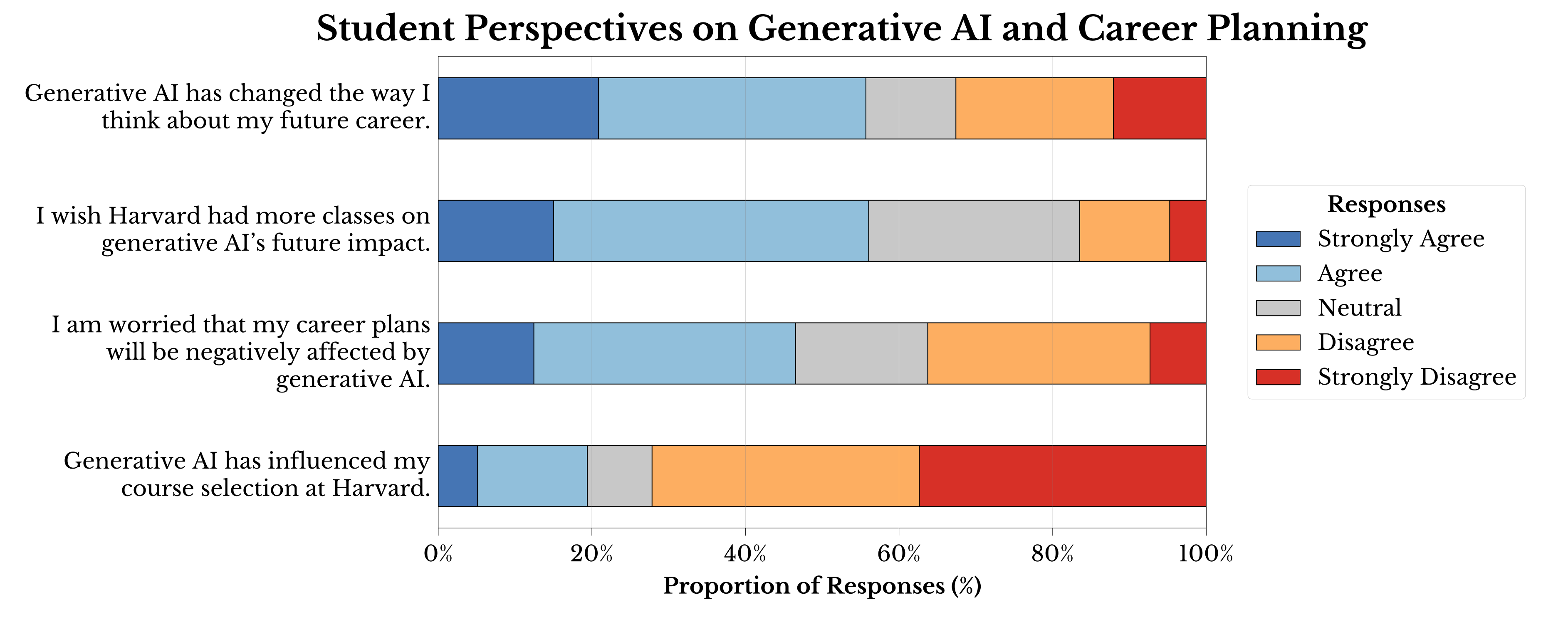}
    \caption{Impact of AI on students' coursework and career plans. These questions were asked to all respondents.}
    \label{fig:courses}
\end{figure}

Figure \ref{fig:courses} shows how generative AI has affected students' perspectives on their coursework and career prospects. Notably, 20\% of students report that AI has influenced their course selection, and around 55\% wish that Harvard had more classes that covered the future impact of generative AI.

\begin{figure}
    \centering
    \includegraphics[width=\linewidth]{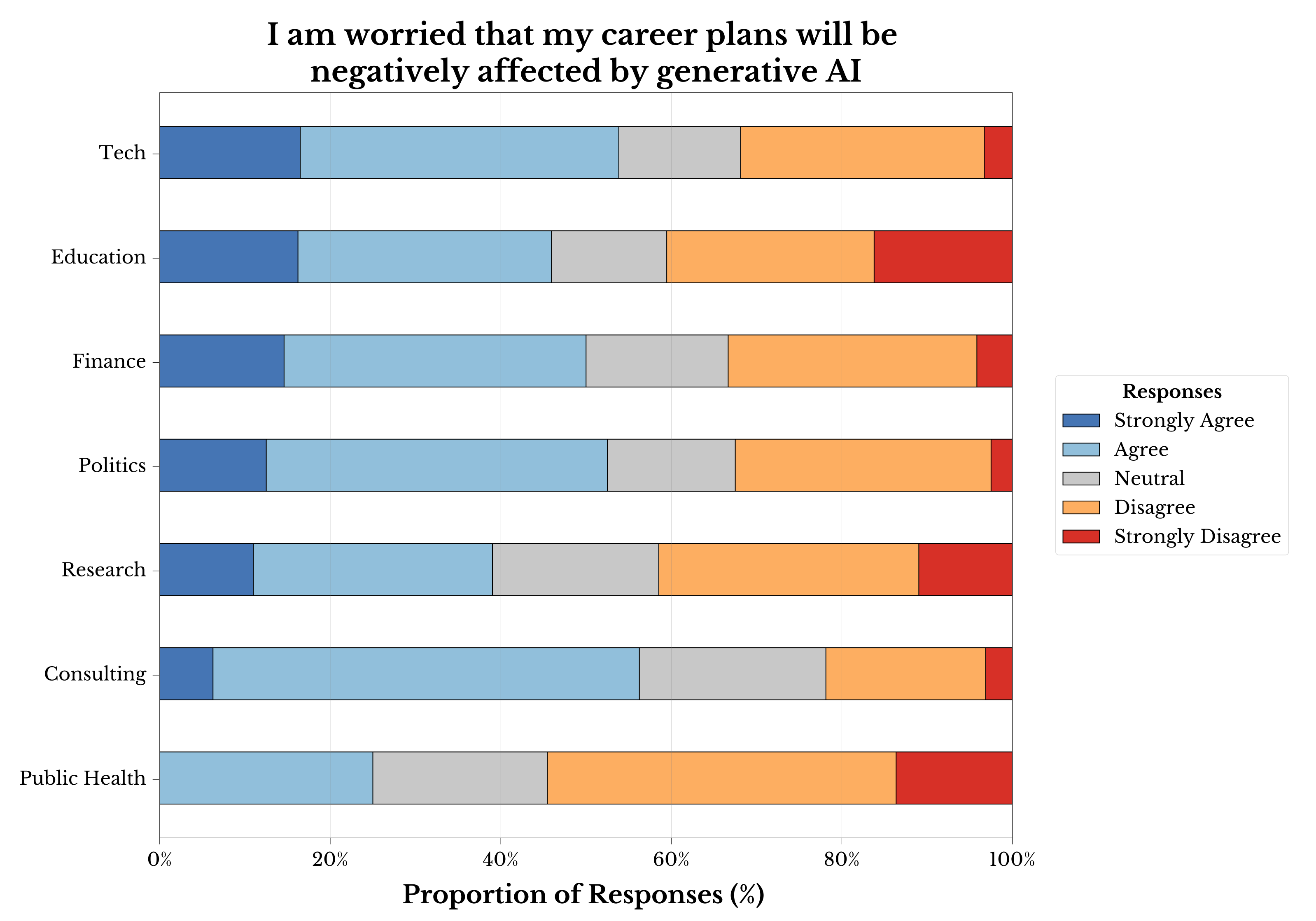}
    \caption{Students' fears of AI negatively affecting their career plans, broken down by career direction. A single respondent may be included in multiple categories if they indicated interest in more than one career direction. \textit{Sample sizes are as follows. Tech: 91, Research: 82, Finance: 48, Public Health: 44, Politics: 40, Education: 37, Consulting: 32.}}
    \label{fig:worries_by_career}
\end{figure}

Generative AI has also changed the way that 55\% of students think about their future careers, and around 45\% of students are worried that AI will negatively affect their career plans. Interestingly, this proportion is roughly the same across almost all categories of career aspirations (Figure \ref{fig:worries_by_career}). Concerns about job loss are paralleled by answers to the free-response question summarized in Table \ref{tab:categorized_worries}.

\subsection{Broader societal concerns} \label{sec:broader_concerns}

\begin{figure}
    \centering
    \includegraphics[width=\linewidth]{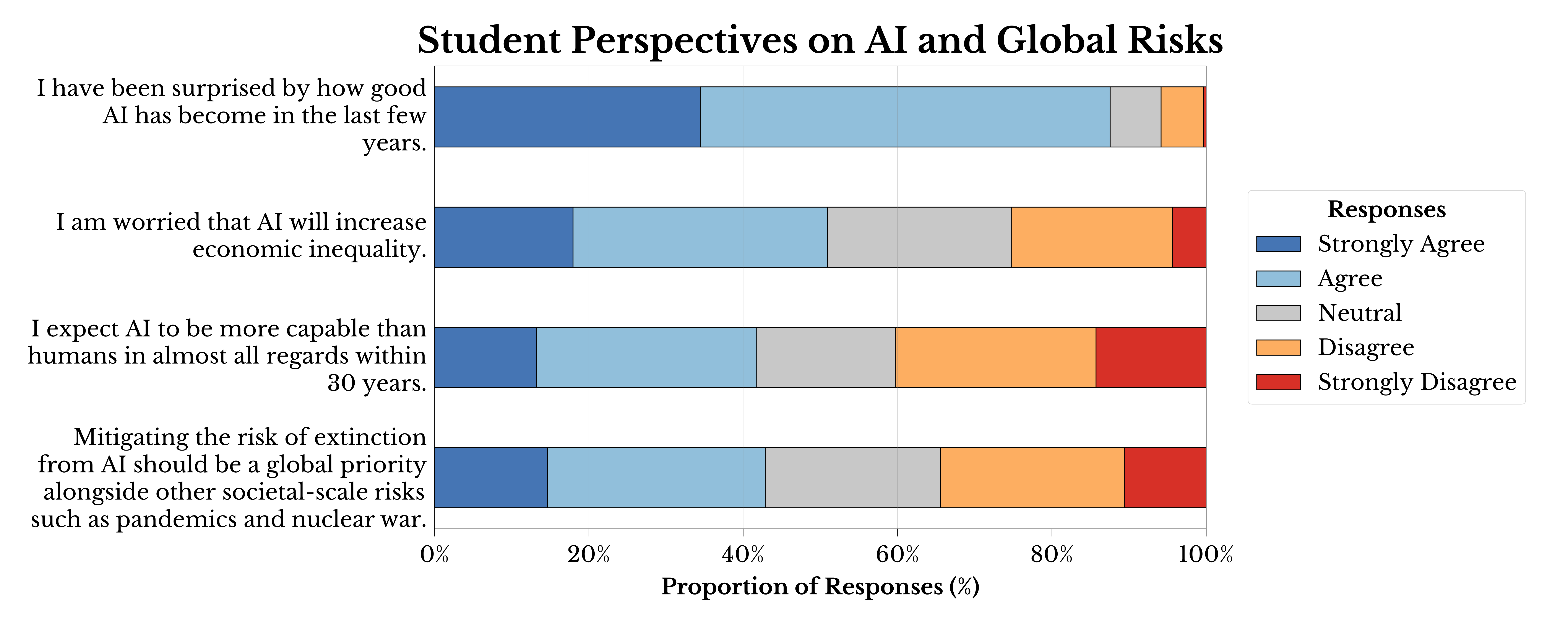}
    \caption{Students' perspectives on global risks from AI. Almost all students were surprised by the rate of AI progress, and many are worried about large-scale economic disruption or human extinction from AI.}
    \label{fig:risks_and_worries}
\end{figure}

The last section of the survey covered the implications of AI for society at large, including risks AI poses to economic equality and human extinction. The main results are summarized in Figure \ref{fig:risks_and_worries}. Most notably, almost all (85\%) students have been surprised by the speed of AI progress in recent years, and around 40\% of students believe that AI systems will be more capable than humans in almost all regards within 30 years.

\begin{figure}
    \centering
    \includegraphics[width=\linewidth]{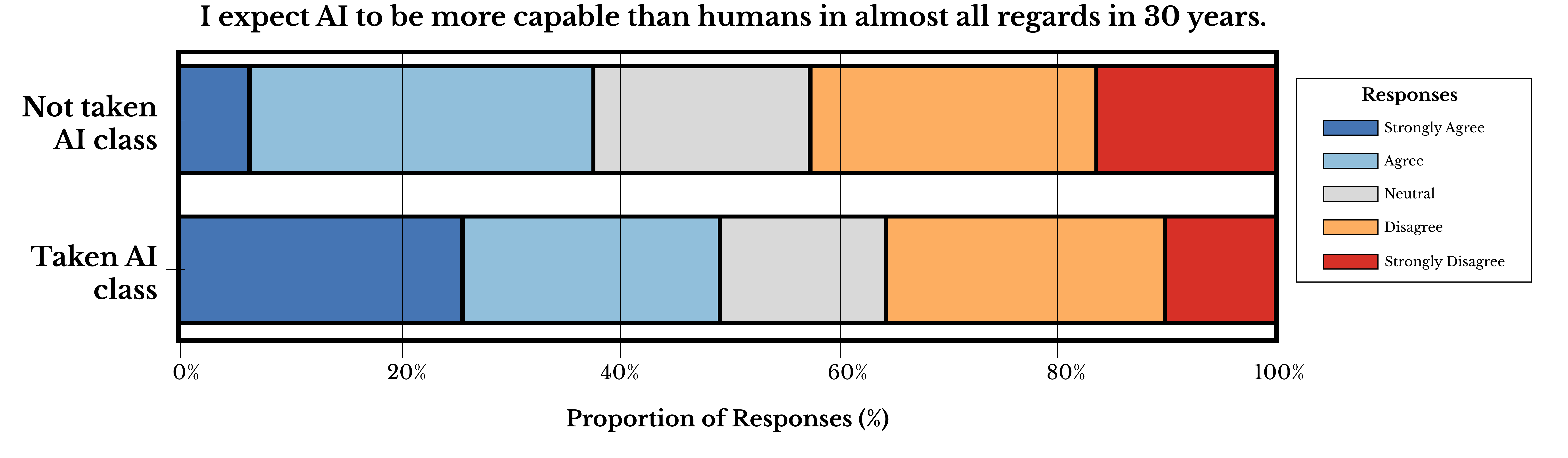}
    \includegraphics[width=\linewidth]{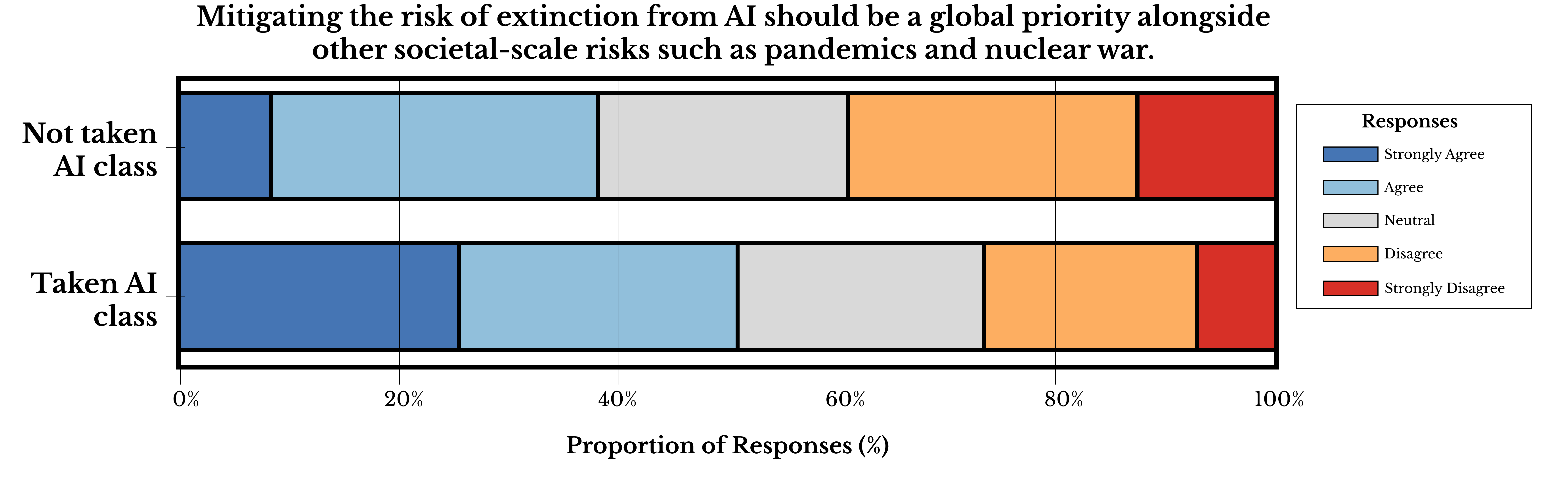}
    \caption{Students respondents who have taken a computer science class on AI tended to have shorter timelines of AI exceeding human capabilities and were more likely to agree that extinction risk from AI should be taken seriously. The expectations of students who have taken an AI class roughly line up with the expectations of AI experts, who on average expect AI to achieve human-level capabilities on all tasks by 2047 with 50\% probability \cite{grace2024thousands}. \textit{Sample size of 98 for Taken AI class, 175 for Not taken AI class.}}
    \label{fig:xrisk_by_experience}
\end{figure}

Relatedly, around 40\% of students agree with the exact wording of a public statement recently put out by the Center for AI Safety (CAIS): ``Mitigating the risk of extinction from AI should be a global priority alongside other societal-scale risks such as pandemics and nuclear war'' \cite{cais}. Since its release in 2023, the CAIS statement has been signed by hundreds of AI scientists and public figures. Students who have taken a computer science class in AI are more likely to agree with the CAIS statement (Figure \ref{fig:xrisk_by_experience}), with a chi-squared test indicating statistically significant differences ($p = 0.024$ and $p = 0.005$ for the timeline and risk of extinction questions, respectively).

\begin{table}[H]
    \centering
    \begin{tabular}{|c|c|p{3in}|}
    \hline
    Category & Count & Example response \\
    \hline
    Job loss & 44 & {\footnotesize ``How many people will be replaced by AI, and how much worse the world will be for it''} \\ \hline
    Overreliance & 17 & {\footnotesize ``degrading our minds and making original problem solving and creativity harder.''}\\ \hline
    Misinformation & 16 & {\footnotesize ``I worry that generative AI will flood our data ecosystem with much more data, that is inherently false. For example, use of genAI for the creation of articles.''} \\ \hline
    Bad actors / Misuse & 13 & {\footnotesize ``The most pressing issue is probably the societal risk of AI weaponry.''} \\ \hline
    Inequality & 11 & {\footnotesize ``Generative ai does put a lot of the conventional jobs out of business, and so countries and regions who have not caught up in terms of the relevant skill capital for the future will lag even more behind.''}\\ \hline
    Extinction & 9 & {\footnotesize ``I'm worried about near-future, human-level AI systems doing AI R\&D, leading to wildly superhuman AI systems that take over the world - either via a human directing them to do so, or via their own "volition."''}\\ \hline
    Bias & 3 & {\footnotesize ``It's incredibly biased.''} \\ \hline
    Miscellaneous & 21 & \\ \hline
    \end{tabular}
    \caption{Answers to the free-response question ``What are your biggest concerns about generative AI in the future?'', categorized by the nature of the worry. Categorization was done by hand.}
    \label{tab:categorized_worries}
\end{table}

\begin{figure}[H]
    \centering
    \includegraphics[width=\linewidth]{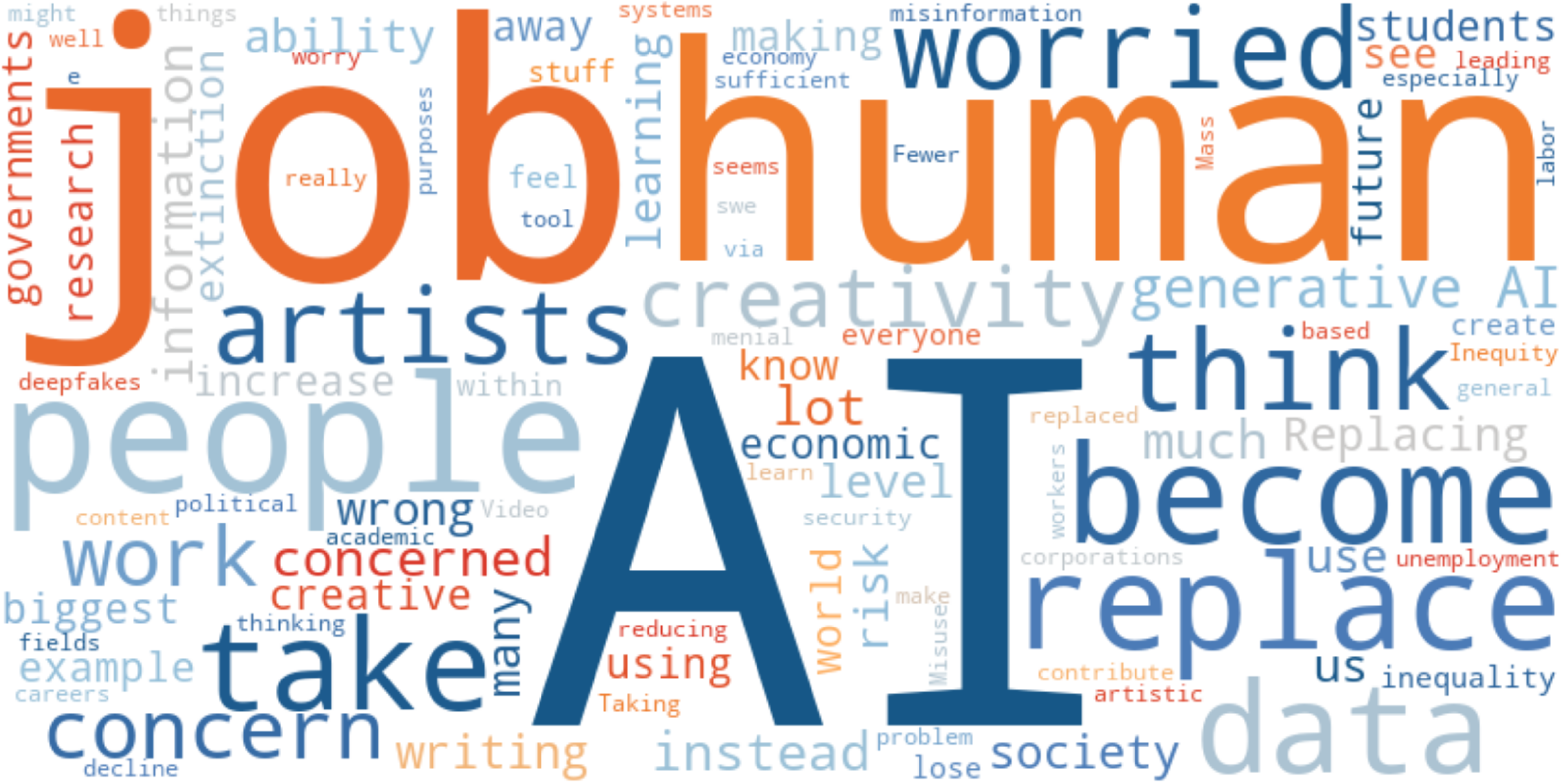}
    \caption{A word cloud of the responses to the free-response question ``What are your biggest concerns about generative AI in the future?''.}
    \label{fig:wordcloud_worries}
\end{figure}

\section{Recommendations}\label{sec:recommendations}

In this section, we present some recommendations to the Harvard administration that are informed by the results of the survey. These include recommendations for course offerings, mental health resources, career support, and more generally, preparation for societal-scale disruption from AI. While these recommendations are made with Harvard in mind, they also apply to other universities.

First, Harvard can take several low-cost, practical steps to help students learn more effectively from AI:
\begin{itemize}
    \item \textbf{Facilitating access to AI:} Appendix \ref{app:spending} demonstrates that some students are getting more value out of AI than others, in large part due to paying for premium AI subscriptions. Harvard should consider providing its students with free access to a paid plan of ChatGPT or Claude.\footnote{Harvard currently provides students access to an AI sandbox that includes GPT-4, but very few students know about this.}
    \item \textbf{Establishing and enforcing consistent rules on AI use:} 30\% of students are worried that their peers are using generative AI to gain an unfair advantage in class. This can be mitigated by promoting assignments that explicitly allow generative AI use. However, this may be infeasible for certain types of assignments like exams based on factual knowledge. Professors may have to increasingly administer these assignments in an in-person setting where a no-AI policy can be easily enforced (or, abandon them altogether).
    \item \textbf{Providing AI-aware career planning services:} To help students more thoughtfully consider the negative impacts of recent and future automation on careers, we suggest that the Mignone Center for Career Services offer AI-aware career planning services. An example event could be a panel by Harvard alumni who have first-hand experience with the impact of AI on the job market.
\end{itemize}

The transformative potential of AI is leaving many students uncertain about how to plan for their future. Harvard should help students navigate their uncertainties and anxieties:

\begin{itemize}
    \item \textbf{Offering courses exploring the future impacts of AI:} There are currently very few classes at Harvard exploring the implications of increasingly powerful AI systems on society, the economy, and the pace of technological progress.\footnote{Three such examples are the new spring 2024 courses \textit{GENED 1188: Rise of the Machines? Understanding and Using Generative AI},  \textit{GOV 94OL: Artificial Intelligence: Sociolegal Dilemmas and Policy Design}, and \textit{COMPSCI  90NDR: Case Studies in Public and Private Policy Challenges of Artificial Intelligence}.} Over half of survey respondents wish there were more of these classes. Some examples of course topics are: How can we model the impact of increased automation on the labor market and economic equality? What will be the influence of AI on democratic decision-making? How can we mitigate the biggest social and existential risks from AI? Harvard could consider hiring new faculty to teach these courses if there is a lack of current faculty with the relevant expertise or interest.
    \item \textbf{Helping students find meaning in education and beyond:} 40\% of students expect AIs to outperform humans within their lifetimes. Students may increasingly struggle to find meaning in their Harvard education as AI systems begin to demonstrate skills beyond those of humans in many domains, and AI progress might cast doubt on the certainty and impact of their future plans. We suggest the establishment of a philosophically-inclined GENED course on finding meaning in an increasingly automated world, so that students can begin to imagine what a prosperous society might look like after human-level AI.
    \item \textbf{Providing mental health support:} We recommend that Harvard provide mental health resources and support groups for students to reflect on stressful questions raised by AI. Harvard should also anticipate possible mental health crises from increased automation, loss of meaning, or potential AI-caused catastrophes. We also suggest that the University establish student advisory groups to inform constantly-evolving university AI policies.
\end{itemize}

\section{Conclusion}\label{Conclusion}

Almost all of us have been surprised by the pace of recent progress in generative AI technology, and it is plausible that this trend will continue for many years. We find that, already, AI is changing the way students consume information, study for classes, and think about their careers.

As a global leader in higher education and research, Harvard University should invest more resources into both 1) adapting to the current impacts of AI on education and 2) anticipating new sources of disruption that may be caused by future AI development.

We hope that this report paves the way for future surveys, at Harvard and beyond, that study the impact of AI on student experiences.


\part*{References}
\addcontentsline{toc}{part}{References}
\bibliography{main.bib}

\newpage
\appendix
\renewcommand{\thesection}{\Alph{section}}
\renewcommand{\thesubsection}{\roman{subsection}}
\renewcommand{\theequation}{A-\arabic{equation}}

\part*{Appendix}
\addcontentsline{toc}{part}{Appendix}


\section{Demographic information} \label{app:representativeness}

How well do our survey respondents reflect the overall Harvard undergraduate population? We break down this question into three salient demographic features: class year, gender, race, and primary area of study.

Compared to overall Harvard undergraduate demographic data taken from \cite{factbook} and \cite{commondataset}, chi-squared tests find that our survey respondents are representative of class year ($p = .43$) and gender ($p = .48$), but are \textit{not} representative of race ($p = .002$) and concentration ($p = 1.9\times 10^{-14}$).\footnote{Gender, race, and concentration data were grouped into the categories seen in the figures. Only respondents who passed the reading comprehension test were included. Some categories, like Native American or Alaska Native, were excluded from this analysis due to small sample sizes. Joint and double concentrations were recorded as a half-count in each subject's category.}

\begin{figure}[h]
    \centering
    \includegraphics[width=.8\linewidth]{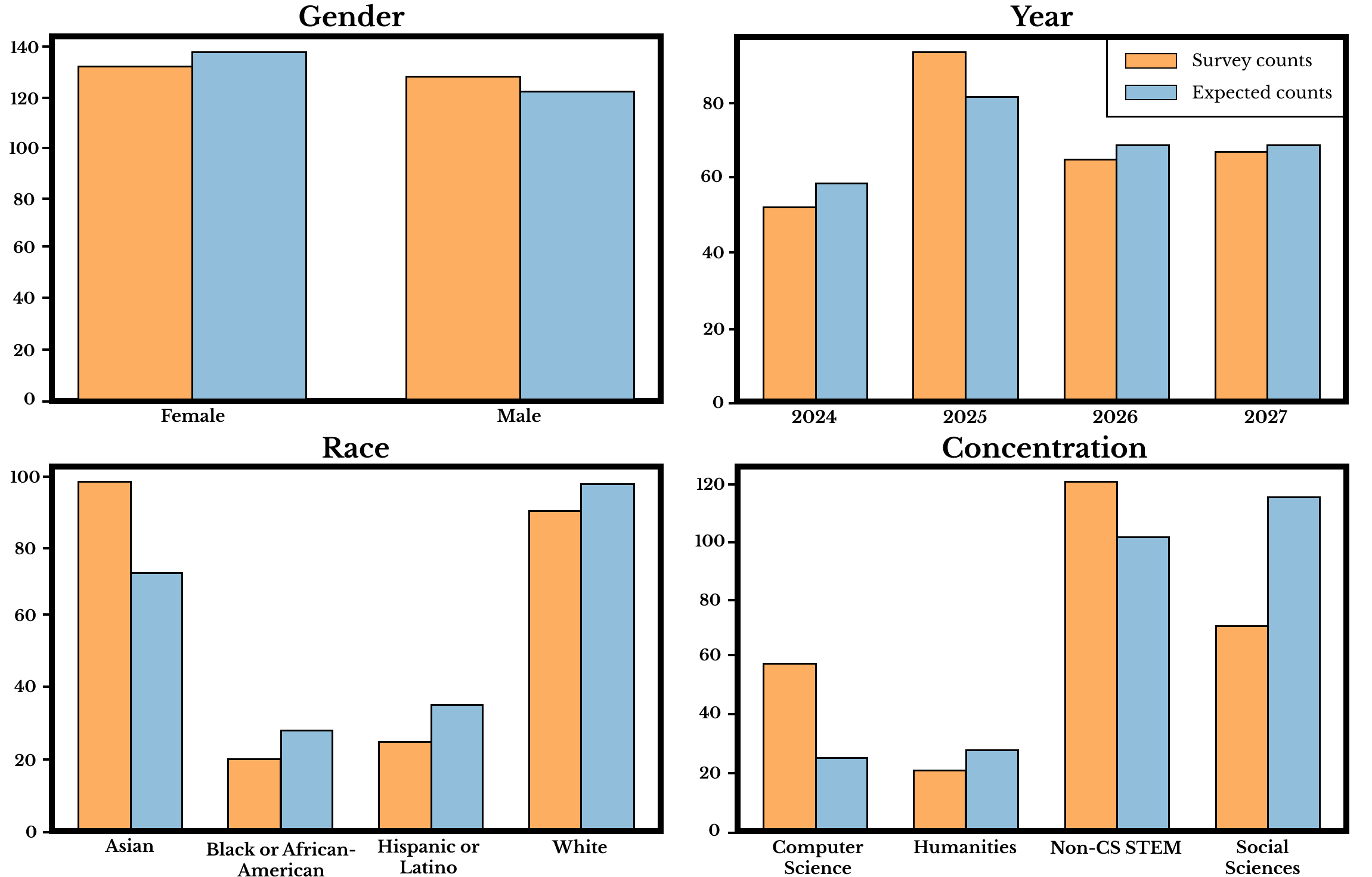}
    \caption{Demographics of survey respondents, compared to the expected demographics if respondents were perfectly representative of the Harvard undergraduate student body.}
    \label{fig:demographics}
\end{figure}

Figure \ref{fig:demographics} compares the true demographics of survey respondents with population averages of Harvard undergrads. Asians, computer science, and STEM concentrators are overrepresented, while social science concentrators are underrepresented. This is to be expected --- students interested in technology are more likely to open a survey about AI. However, our sample still appears reasonably proportional overall, so we do not believe this significantly affects the results of the survey.

\section{Impact of paid AI subscriptions} \label{app:spending}

\begin{figure}[h]
    \centering
    \includegraphics[width=.5\linewidth]{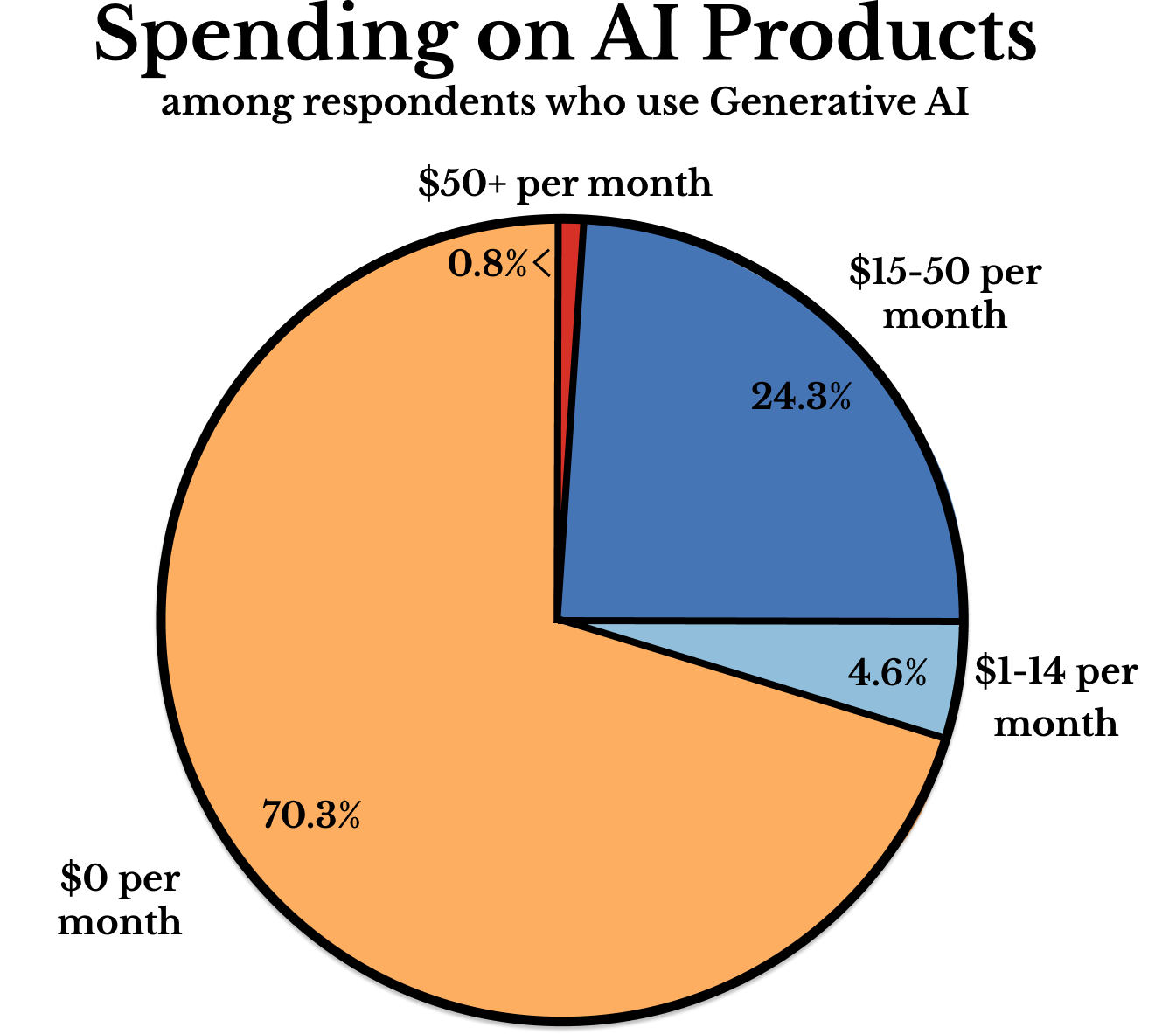}
    \caption{How much money students spend on AI products per month, among respondents who use generative AI. We hypothesize that the majority of people in the \$15-\$50 category were subscribed to ChatGPT Pro, which costs \$20 per month.}
    \label{fig:spending}
\end{figure}

Figure \ref{fig:spending} shows that 30\% of students who use generative AI pay for premium subscriptions, usually ranging from \$15 to \$50 per month. How much does this increase the value that students get out of AI?

\begin{figure}[H]
    \centering
    \includegraphics[width=.7\linewidth]{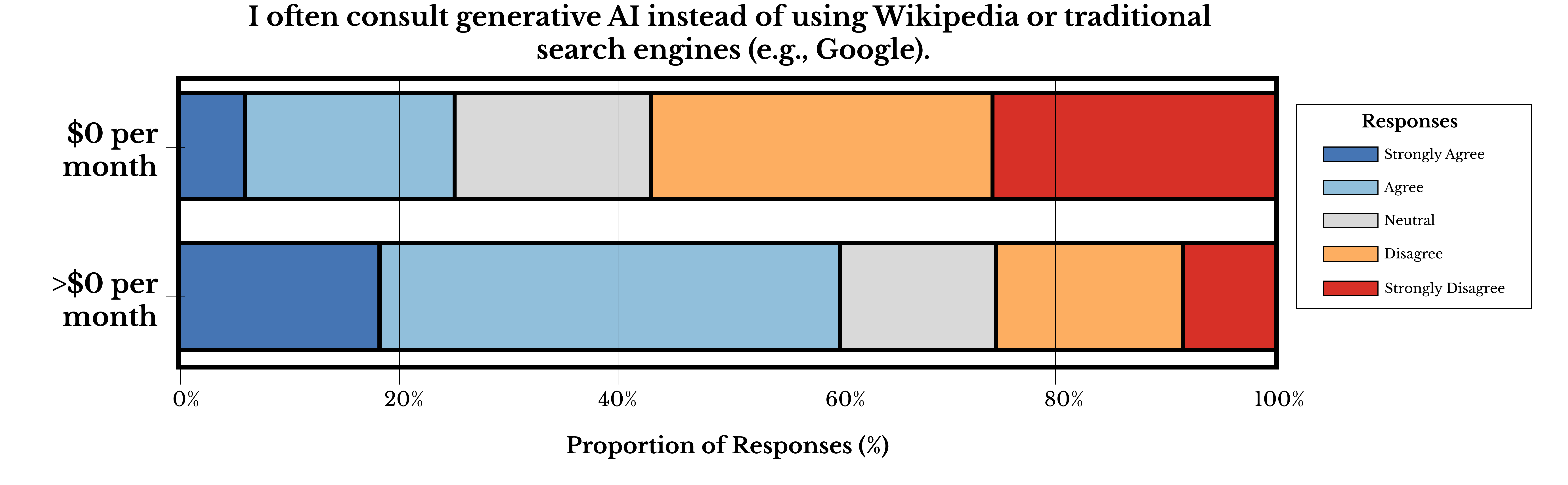}
    \includegraphics[width=.7\linewidth]{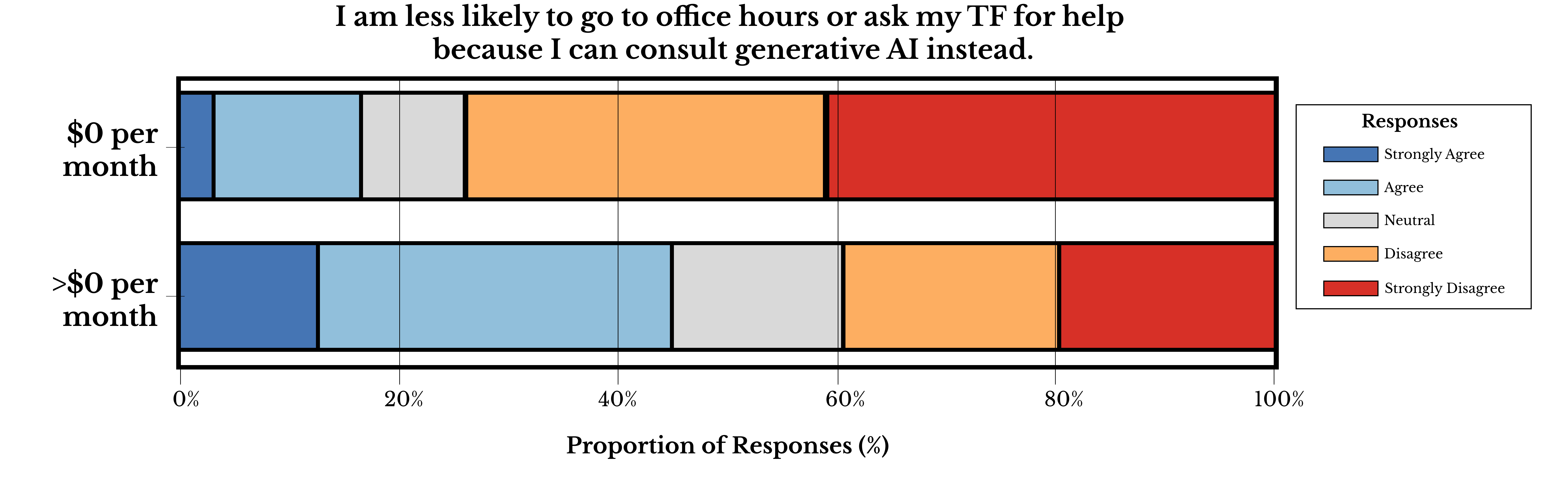}
    \includegraphics[width=.7\linewidth]{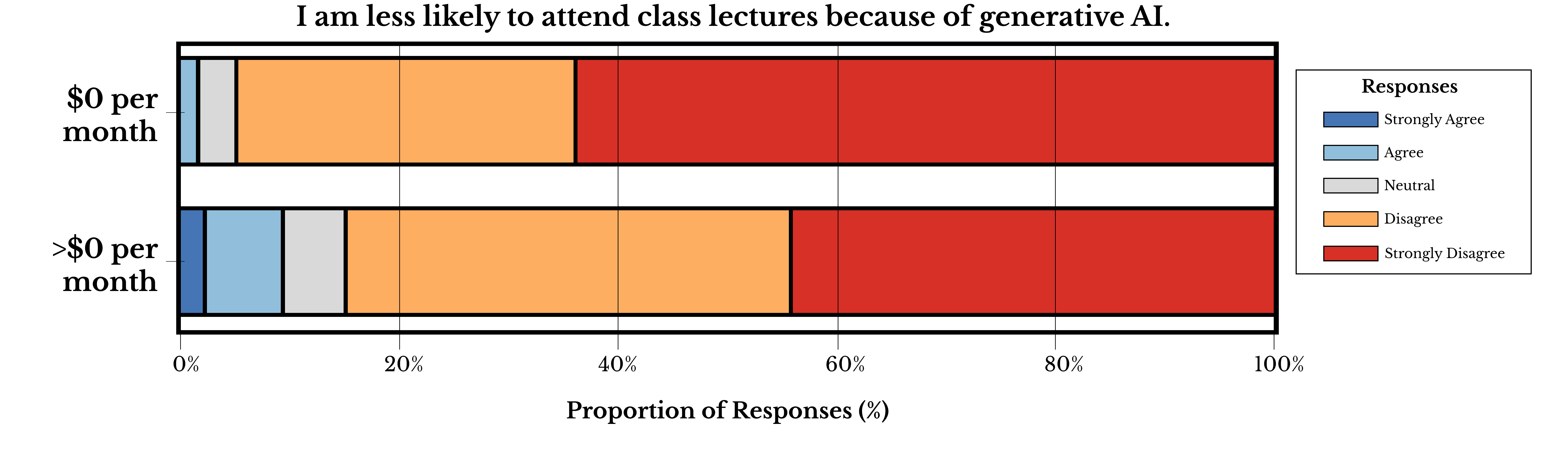}
    \caption{Students who spend money on AI report getting more use out of it and rely less on traditional resources. \textit{Sample size of 168 for \$0 per month, 71 for >\$0 per month.}}
    \label{fig:habits_by_spending}
\end{figure}

The answer seems to be: quite a lot. Figure  \ref{fig:habits_by_spending} shows that, compared to students who only use free AI products, students who pay for AI subscriptions are over twice as likely to use generative AI products instead of Wikipedia or Google search. They are also almost three times as likely to report decreased utilization of office hours because they can consult AI instead. Chi-squared tests illustrate statistically significant differences between the free and paid generative AI product groups for all three questions: $p = 9.49 \times 10^{-6}$, $p = 3.54 \times 10^{-5}$, and $p = 0.0214$ for consulting generative AI instead of Wikipedia or Google, less likely to go to office hours, and less likely to attend class lectures, respectively.

The causal direction here is not obvious. It may be that students who use AI more frequently are more willing to pay for premium AI products. Or, it may be that students with premium AI products consistently get higher-quality outputs, which encourages them to use AI more often (GPT-4, which at the time of the survey was only accessible through a paid subscription, is known to be significantly more useful than GPT-3.5 \cite{bubeck2023sparks}). The truth likely lies somewhere in the middle.

\begin{figure}[h]
    \centering
    \includegraphics[width=.7\linewidth]{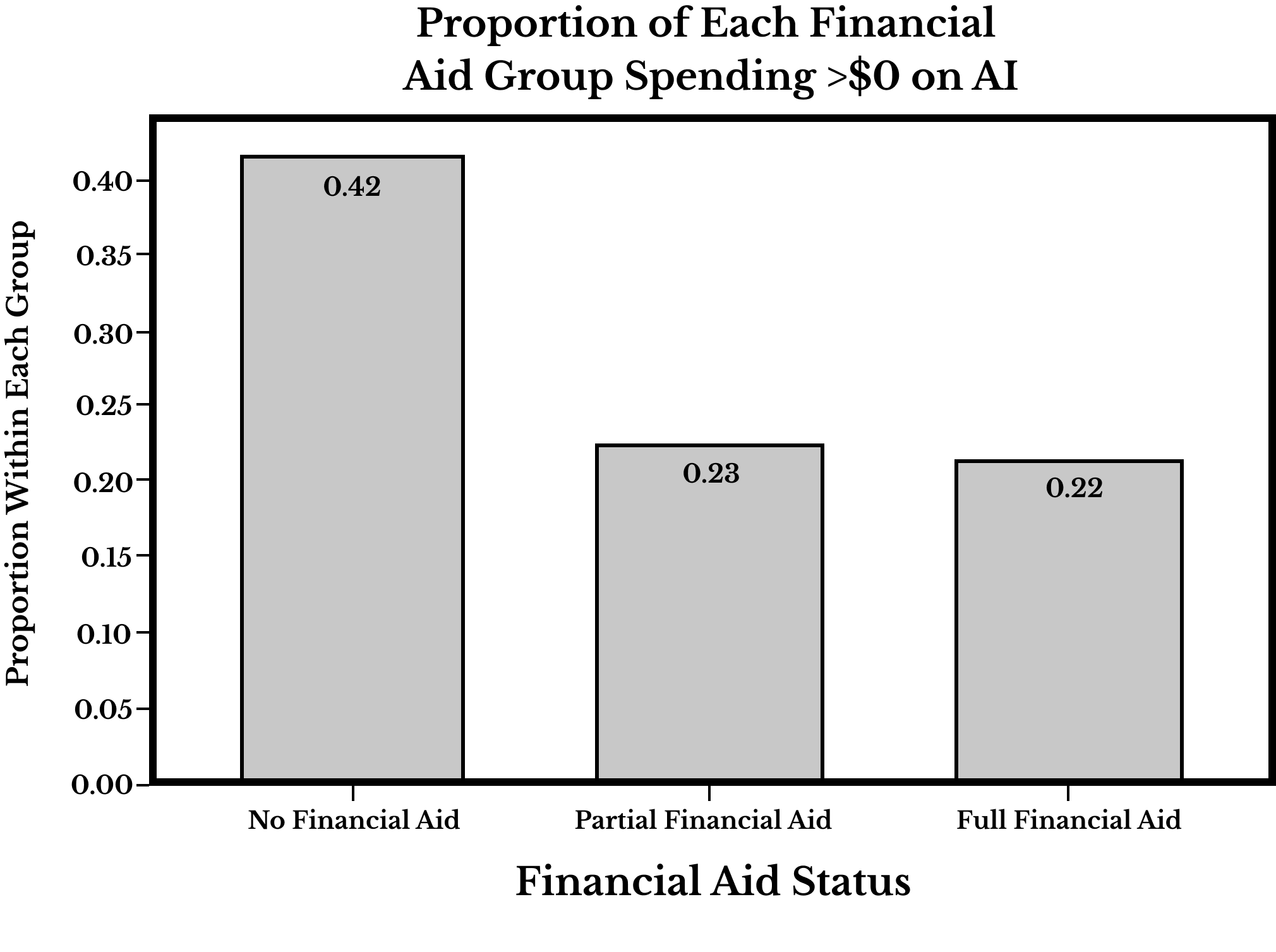}
    \caption{Students who don't receive financial aid were twice as likely to pay for subscription AI products than students who do receive financial aid. \textit{Sample size of 79 for No financial aid, 62 for Partial financial aid, and 60 for Full financial aid.}}
    \label{fig:spending_by_finaid}
\end{figure}

Finally, we investigate how spending on AI products is correlated with socioeconomic background. We find that, among AI users, 40\% of students who receive no financial aid pay for AI products, compared to 20\% of students with partial or full financial aid (Figure \ref{fig:spending_by_finaid}). To the extent that 1) this difference is due to the cost of AI being prohibitive instead of other correlated factors, and 2) premium AI subscriptions actually improve student experiences, this is a potentially concerning source of student inequity.

\pagebreak
A chi-squared test comparing the proportions of these three different groups yields statistically significant differences ($p = 0.012$), indicating that there is an association between financial aid status and spending on AI products.

\section{Additional figures} \label{app:more_figures}

\begin{figure}[H]
    \centering
    \includegraphics[width=\linewidth]{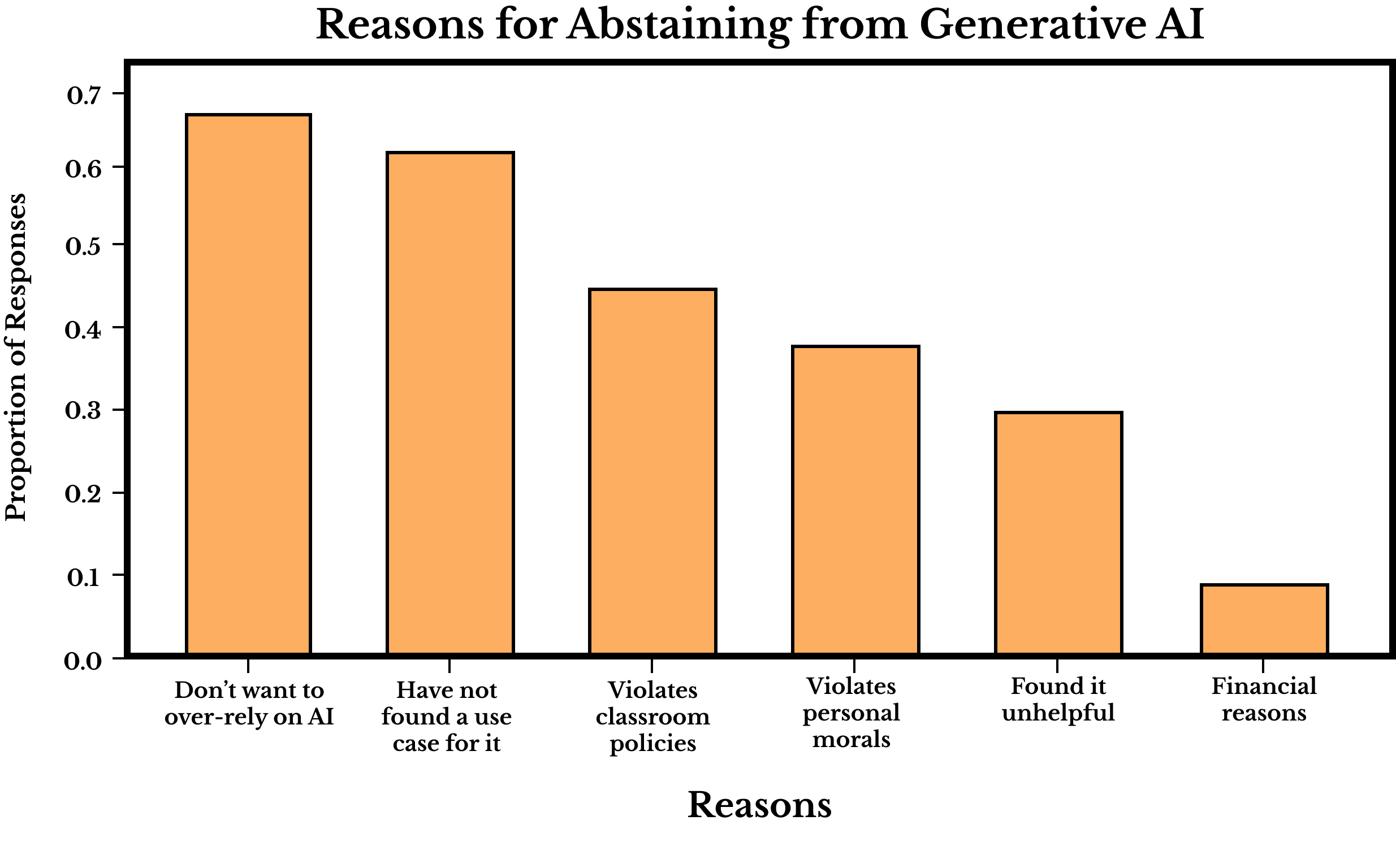}
    \caption{Students who reported not using generative AI cited a variety of reasons for their abstinence. The most common reasons are not wanting to develop a reliance on AI and not having found a good use-case for it. \textit{Sample size of 34.}}
    \label{fig:reasons_abstain}
\end{figure}

\begin{figure}[H]
    \centering
    \includegraphics[width=\linewidth]{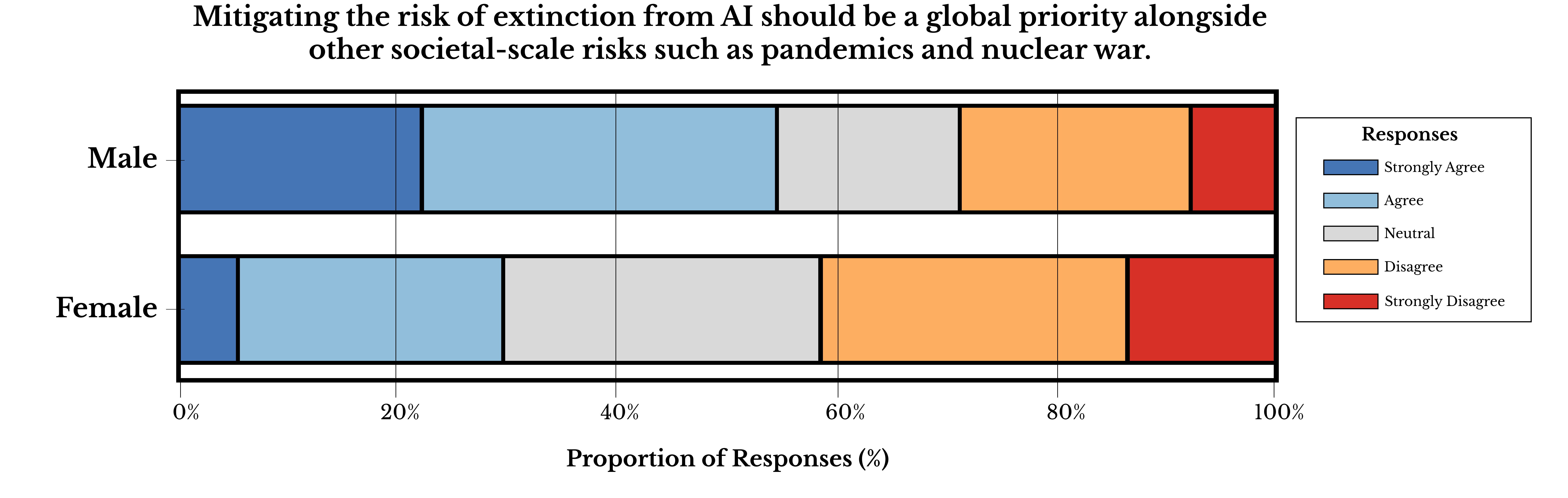}
    \includegraphics[width=\linewidth]{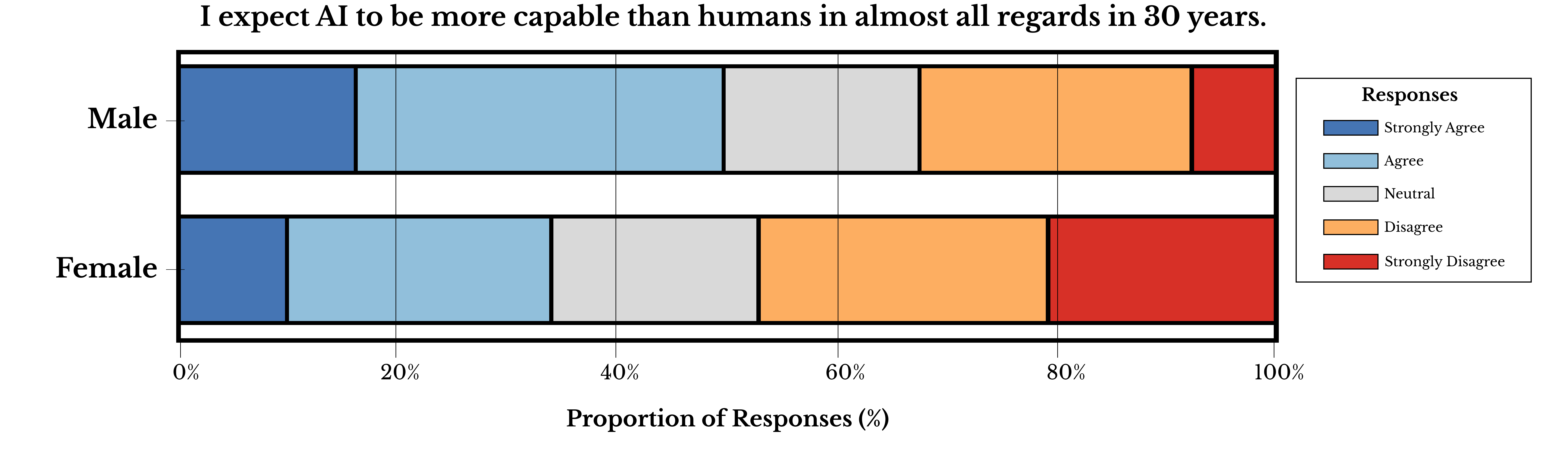}
    \caption{Among Harvard undergraduates, male students tended to have shorter AI timelines than females and were more likely to be concerned about the risk of extinction from AI. \textit{Sample size of 129 for Male, 133 for Female.}}
    \label{fig:cais_timelines_gender}
\end{figure}

For the statement, ``Mitigating the risk of extinction from AI should be a global priority alongside other societal-scale risks such as pandemics and nuclear war,'' a chi-squared test gives $p = 0.053$, slightly above the 0.05 significance threshold. However, for the statement, ``I expect AI to be more capable than humans in almost all regards in 30 years,'' $p = 0.002$, indicating a significant difference between male and female responses.

\begin{figure}[H]
    \centering
    \includegraphics[width=\linewidth]{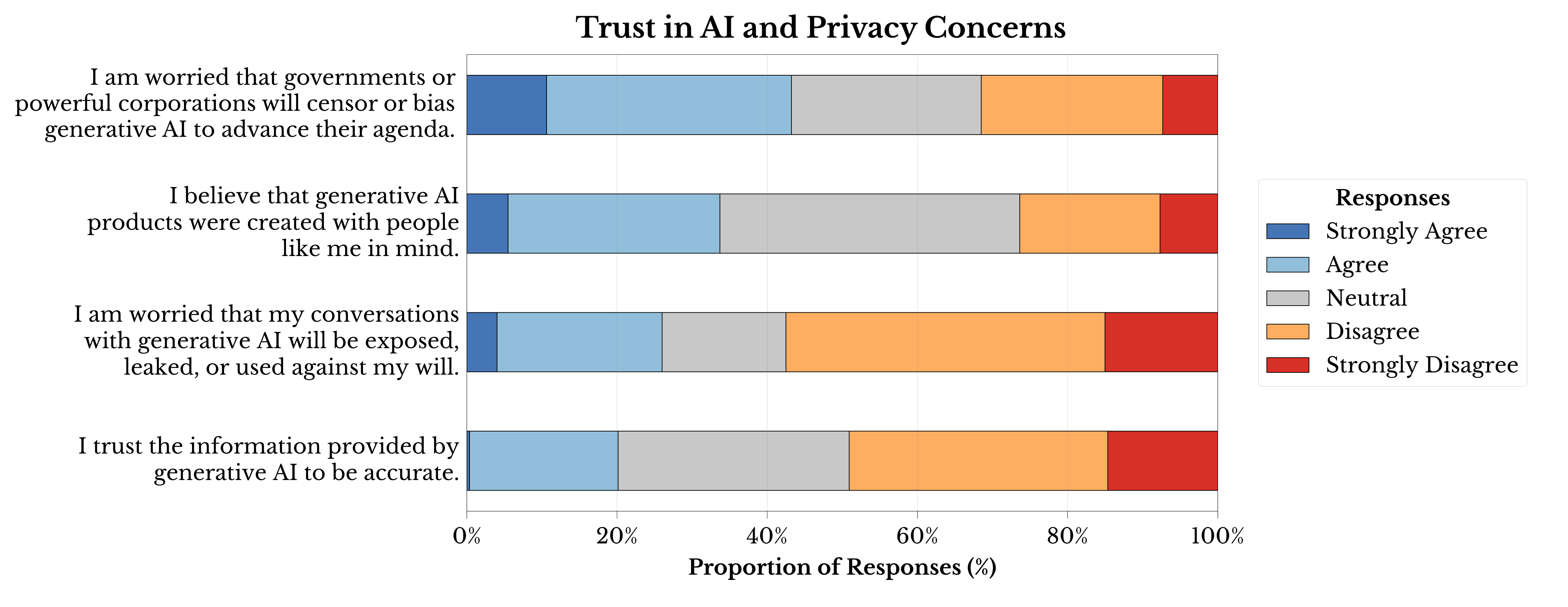}
    \caption{Students are generally skeptical of the accuracy of the outputs of generative AI. Further, 40\% of students are worried that corporations will censor or bias generative AI to advance their agenda.}
    \label{fig:trust_in_ai}
\end{figure}

\begin{figure}[H]
    \centering
    \includegraphics[width=\linewidth]{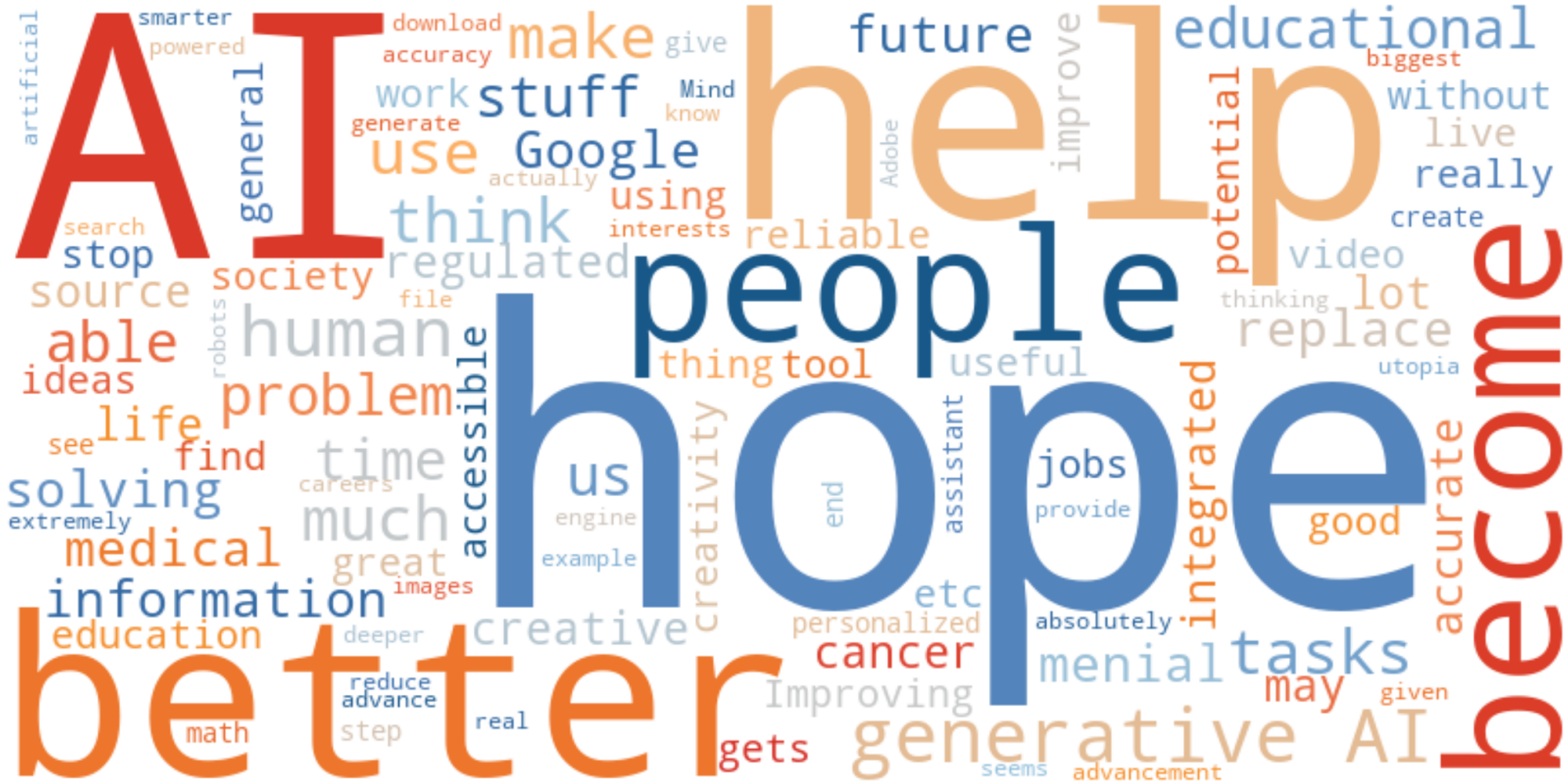}
    \caption{A word cloud of the responses to the free-response question ``What are your biggest hopes for generative AI in the future?''}
    \label{fig:wordcloud_hopes}
\end{figure}

\section{Survey details} \label{app:survey_details}

The survey accepted responses from April 18 to April 24, 2024. It was publicized to all Harvard undergraduates in a Harvard Undergraduate Association email update and undergraduate house mailing lists. To encourage students to complete the survey, 100 random respondents were given \$10 gift cards. All responses were collected anonymously.

One of the questions on the survey was ``While using generative AI, I have been abducted by an alien. Choose the second option from the left/top.'' All participants who did not choose ``Disagree'' (which corresponded to the second option from the left) had their responses excluded from our analysis. We did this to make sure that we only included data from students who were reading the questions carefully.

All of the multiple-choice questions were required, except for the demographics questions about financial aid and career plans. The free response questions were all optional.

\section{All survey questions}
Here is the exact text of all questions on the survey.

\begin{itemize}
    \item What is your (intended) concentration? If pursuing a joint or double concentration, select both. Use command-click to select multiple if you are on a computer. \textit{All concentrations were provided as options. Respondents could select up to two answers.}
    \item If you have a secondary, please indicate so here. \textit{All secondaries were provided as options.}
    \item Which graduating class are you a part of?
    \begin{itemize}
        \item \textit{Class of 2024; Class of 2025; Class of 2026; Class of 2027}
    \end{itemize}
    \item What is your gender?
    \begin{itemize}
        \item \textit{Male; Female; Other; Prefer not to say}
    \end{itemize}
    \item What is your race/ethnicity? \textit{Respondents could select multiple answers.}
    \begin{itemize}
        \item \textit{White; Black or African-American; Native American or Alaska Native; Asian; Native Hawaiian or Pacific Islander; Hispanic or Latino; Other; Prefer not to say}
    \end{itemize}
    \item Do you receive need-based financial aid? (Optional)
    \begin{itemize}
        \item \textit{No financial aid; Partial financial aid; Full financial aid; Prefer not to say}
    \end{itemize}
    \item Which areas best describe your career plans? (Optional) \textit{Respondents could select multiple answers.}
    \begin{itemize}
        \item \textit{Tech; Politics; Consulting; Finance; Public Health; Research; Education; Other [please specify]}
    \end{itemize}
    \item Have you ever taken a computer science class on AI or Machine Learning? \textit{Yes; No}
    \item Do you ever use generative AI products? These include chatbots, image generators, or AI music generators. \textit{Yes; No}
\end{itemize}
The following questions were only shown to respondents who answered ``Yes'' to the question about ever using generative AI.
\begin{itemize}
    \item How often do you use generative AI chatbots (like ChatGPT, Claude, Gemini, Perplexity AI, etc.)?
    \begin{itemize}
        \item \textit{Rarely Ever; Biweekly; Weekly; Every Other Day; Daily / Almost Daily}
    \end{itemize}
    \item Which of the following generative AI products do you use? \textit{Respondents could select multiple answers.}
    \begin{itemize}
        \item \textit{ChatGPT; Claude; Gemini; Perplexity AI; GitHub Copilot; Midjourney; Other [please specify]}
    \end{itemize}
    \item How much do you spend per month on generative AI products? (Note that ChatGPT Pro [GPT-4] costs \$20 per month.)
    \begin{itemize}
        \item \textit{\$0 per month; \$1-14 per month; \$15-50 per month; Over \$50 per month}
    \end{itemize}
    \item What do you use generative AI for? \textit{Respondents could select multiple answers.}
    \begin{itemize}
        \item \textit{Programming assignments; Data processing; Writing assignments (coming up with ideas, drafting, proof-reading); Writing emails; To answer general questions (``How does a 401k work?''); Entertainment or companionship; Creating graphics, art, or other creative work; Translation or language learning; Other [please specify]}
    \end{itemize}
    \item I often consult generative AI instead of using Wikipedia or traditional search engines (e.g., Google).
    \begin{itemize}
        \item \textit{Strongly Disagree; Disagree; Neutral; Agree; Strongly Agree}
    \end{itemize}
    \item I am less likely to do the readings for class since I can ask generative AI to summarize the readings.
    \begin{itemize}
        \item \textit{Strongly Disagree; Disagree; Neutral; Agree; Strongly Agree}
    \end{itemize}
    \item I am less likely to go to office hours or ask my TF for help because I can consult generative AI instead.
    \begin{itemize}
        \item \textit{Strongly Disagree; Disagree; Neutral; Agree; Strongly Agree}
    \end{itemize}
    \item I am less likely to attend class lectures because of generative AI.
    \begin{itemize}
        \item \textit{Strongly Disagree; Disagree; Neutral; Agree; Strongly Agree}
    \end{itemize}
\end{itemize}

The following question was only shown to respondents who answered ``No'' to the question about ever using generative AI.
\begin{itemize}
    \item What are the reasons you abstain from using generative AI? \textit{Respondents could select multiple answers.}
    \begin{itemize}
        \item \textit{Financial reasons; Have not found a use case for it; Found it unhelpful; Don't want to over-rely on AI; Violates classroom policies; Violates personal morals; Other [please specify]}
    \end{itemize}
\end{itemize}
The following questions were multiple choice, with the options Strongly Disagree, Disagree, Neutral, Agree, and Strongly Agree.
\begin{itemize}
    \item I understand the rules regarding the use of generative AI in my classes.
    \item I am worried that my peers will use generative AI to gain an unfair advantage in classes.
    \item I trust the information provided by generative AI to be accurate.
    \item I am worried that my conversations with generative AI will be exposed, leaked, or used against my will.
    \item I am worried that governments or powerful corporations will censor or bias generative AI to advance their agenda.
    \item I believe that generative AI products were created with people like me in mind.
    \item I have been surprised by how good AI has become in the last few years.
    \item Generative AI has influenced my course selection at Harvard.
    \item Generative AI has changed the way I think about my future career.
    \item I wish Harvard had more classes that taught us how to effectively use generative AI.
    \item I wish Harvard had more classes that explored how generative AI will affect the future.
    \item While using generative AI, I have been abducted by an alien. Choose the second option from the left/top.
    \item I am worried that my career plans will be negatively affected by generative AI.
    \item I am worried that AI will increase economic inequality.
    \item I expect AI to be more capable than humans in almost all regards within 30 years.
    \item Mitigating the risk of extinction from AI should be a global priority alongside other societal-scale risks such as pandemics and nuclear war.
\end{itemize}
The following questions were free-response and optional.
\begin{itemize}
    \item What are the policies on generative AI use in your classes? Feel free to name specific classes or departments.
    \item What have you seen generative AI struggle at?
    \item What is the most unique or creative way you have used generative AI?
    \item What are your biggest hopes for generative AI in the future?
    \item What are your biggest concerns about generative AI in the future?
    \item What classes or resources would you like to see regarding education and news about generative AI?
    \item Is there anything else you have to say or want to share?
    \item Did you experience any problems with this survey?
\end{itemize}
\end{document}